\documentclass[prd,aps,amsmath,amssymb,nofootinbib,preprintnumbers]
{revtex4}  \voffset=1cm
\usepackage{graphicx}
\usepackage{dcolumn}
\usepackage{bm}
\usepackage{amsmath}
\usepackage{amsfonts}
\usepackage{ifthen}
\usepackage{psfrag}
\usepackage{multirow}

\def\ls{\mathrel{\lower4pt\vbox{\lineskip=0pt\baselineskip=0pt
           \hbox{$<$}\hbox{$\sim$}}}}
\def\gs{\mathrel{\lower4pt\vbox{\lineskip=0pt\baselineskip=0pt
           \hbox{$>$}\hbox{$\sim$}}}}
\def\drawbox#1#2{\hrule height#2pt

\hbox{\vrule width#2pt height#1pt \kern#1pt
              \vrule width#2pt}
              \hrule height#2pt}

\def\Asym#1#2{\vcenter{\vbox{\drawbox{#1}{#2}
              \kern-#2pt       
              \drawbox{#1}{#2}}}}

\newcommand{\sss}{\sin^2\!\theta_{12}}
\newcommand{\sts}{\sin^2\!2\theta_{12}}

\newcommand{\css}{\cos^2\!\theta_{12}}

\newcommand\T{\rule{0pt}{3ex}} 

\newcommand{\ul}{\underline}
\newcommand{\mee}{\langle m_{ee}\rangle}
\newcommand{\sumnu}{\sum m_i}

\newcommand{\dms}{\Delta m_{\rm S}^2}
\newcommand{\dma}{\Delta m_{\rm A}^2}
\newcommand{\ratiomue}{\Phi_\mu/\Phi_e}
\newcommand{\ratioetau}{\Phi_e/\Phi_\tau}

\newcommand{\be}{\begin{equation}}
\newcommand{\ee}{\end{equation}}
\newcommand{\bea}{\begin{eqnarray}}
\newcommand{\eea}{\end{eqnarray}}

\begin{document}
\begin{flushright}
UMD-PP-10-023
\end{flushright}
\vspace{0.3in}
\title {\Large Testing the Bimodal/Schizophrenic Neutrino Hypothesis in Neutrinoless
 Double Beta Decay and Neutrino Telescopes}
\author{\bf James Barry$^{1}$\footnote{james.barry@mpi-hd.mpg.de},
Rabindra N.~Mohapatra$^{2}$\footnote{rmohapat@umd.edu} and
Werner Rodejohann$^{1}$\footnote{werner.rodejohann@mpi-hd.mpg.de}}
\affiliation{$^{1}$Max-Planck-Institut f\"ur Kernphysik, Postfach 103980, D-69029 Heidelberg, Germany \\
$^{2}$Maryland Center for Fundamental Physics, Department of Physics, University of Maryland,
 College Park, Maryland 20742, USA}

\begin{abstract}
\noindent

The standard assumption is that all three neutrino mass states are
either Dirac or Majorana. However, it was recently suggested by
Allaverdi, Dutta and one of the authors (R.N.M.) that mixed, or
bimodal, flavor neutrino scenarios are conceivable and are
consistent with all known observations (these were called
``schizophrenic'' in the ADM paper).  In that case each individual
mass eigenstate can be either Dirac or Majorana, so that the
flavor eigenstates are ``large'' admixtures of both. An example of
this ``bimodal'' situation is to consider one mass state as a Dirac particle
(with a sterile partner), while the other two are of Majorana
type. Since only Majorana particles contribute to neutrinoless
double beta decay, the usual dependence of this observable on the
neutrino mass is modified within this scenario. We study this in
detail and, in particular, generalize the idea for all possible
bimodal combinations. Inevitably, radiative corrections will
induce a pseudo-Dirac nature to the Dirac states at the one-loop
level, and the effects of the pseudo-Dirac mass splitting will
show up in the flavor ratios of neutrinos from distant
cosmological sources. Comparison of the effective mass in
neutrinoless double beta decay as well as flavor ratios at
neutrino telescopes, for different pseudo-Dirac cases and with
their usual phenomenology, can distinguish the different bimodal
possibilities.

\end{abstract}

\maketitle

\section{Introduction}

The observation of neutrino masses and mixings provides the first
conclusive evidence for physics beyond the standard model (BSM),
so that an understanding of this phenomenon will open up one clear
direction for the new BSM physics. Knowledge of the nature of the
neutrino mass is crucial in order to make progress in this search.
Unlike quarks and charged leptons, for which the form of the mass
term is unambiguous, neutrinos are electrically neutral and
therefore allow several possibilities. The two kinds of mass terms
widely discussed are: (i)  a Dirac type mass, which requires the
theory to have a lepton number symmetry as well as a right-handed
(sterile) neutrino degree of freedom; or (ii) a Majorana type
mass, which necessitates the breaking of lepton number.
Implementing the first possibility in the standard model requires
a minimal set of assumptions, i.e., simply adding three
right-handed neutrinos. In order to understand the small masses
one requires that the associated Yukawa couplings are of order
$10^{-12}$ or less. The challenge then becomes one of
understanding this tiny Yukawa coupling or at least connecting it
to some other phenomenon that requires it. In the case of Majorana
neutrinos, one can write effective dimension five operators of the
form $LHLH/M$, where $M$ represents the effect of higher scale
physics. While no small couplings need be invoked in this case,
one must understand the origin of the high scale $M$ and explore
what physics is associated with it. The most widely discussed
theories of this type are the seesaw models \cite{seesaw}, where
the higher scale could come from the breaking of new symmetries
such as $B-L$ or possibly a grand unified theory such as $SO(10)$.

An intermediate possibility that has also been discussed in the
literature is the pseudo-Dirac scenario, where a tiny Majorana
mass is added for either one or both of the two two-component
neutrino states that make up the Dirac neutrino \cite{pseudo}. If
one considers all three active neutrinos to be pseudo-Dirac, then
current observations put very stringent constraints on the
magnitude of the Majorana mass \cite{andre}, i.e.,~$\ls 10^{-10}$
eV, in order to have the pseudo-Dirac mass splitting small enough
to remain undetected in solar neutrino oscillation experiments.
Roughly speaking, for $\nu_1$ and $\nu_2$, which contain a large
amount of $\nu_e$, the pseudo-Dirac mass-squared difference should
be smaller than $E/L \sim 10^{-11}$ eV$^{2}$ for solar neutrinos,
otherwise the associated oscillations would have been observed
in solar neutrino data. Detailed explanations can be found in Ref.~\cite{andre}. This tiny splitting makes these neutrinos almost Dirac particles, hence the name pseudo-Dirac.

In a recent paper \cite{adm}, a new possibility for neutrino
masses was pointed out, where some neutrino mass eigenstates are
Dirac while the others are Majorana. This is only
phenomenologically viable if one defines the Dirac or Majorana
nature of neutrinos in terms of the mass eigenstates, rather than
the flavor eigenstates. In this case, all neutrino flavors have
large admixtures of both Dirac and Majorana type mass, and can be
called ``bimodal flavor neutrinos'' (or schizophrenic
neutrinos, as in Ref.~\cite{adm}). One then needs to add as many
sterile neutrino states to the standard model as there are Dirac
mass eigenstates. This is different from the pseudo-Dirac case in
the sense that the lepton number violating and conserving terms
have comparable magnitude. Another interesting feature of bimodal
neutrinos is that unlike the case of pseudo-Dirac flavor
neutrinos, where there exist stringent constraints on the Majorana
admixture, here the oscillations of solar neutrinos (as well as
all other oscillation observations) remain unaffected. In other
words, in conventional neutrino oscillation experiments the
bimodal flavor case looks the same as the pure Dirac or pure
Majorana case.

An obvious place where the bimodal scenario leads to a different
effect from both the pure Majorana and pseudo-Dirac possibilities
is in the predictions for neutrinoless double beta decay. This
was noted for a very specific model in Ref.~\cite{adm}. In this
paper we consider the most general implementation of this idea and
present the predictions for neutrinoless double beta decay for
the cases of both normal and inverted mass ordering.

It was also pointed out in Ref.~\cite{adm} that since there is no
symmetry guaranteeing the bimodal possibility, one-loop
corrections can induce a tiny ($\leq 10^{-14}$ eV) amount of Majorana mass
to the mass eigenstate that had a tree level Dirac mass, effectively making
this mass eigenstate pseudo-Dirac. Although this value is well within the
constraints from solar neutrino observations, there are implications for
 astrophysical neutrinos. This is one of the new points explored in this paper. \\

In Sec.~\ref{sect:models} we provide a brief review of models
that could lead to the bimodal flavor neutrino scenario.
Secs.~\ref{sect:cosmonu} and \ref{sect:0nubb} contain a
discussion of the phenomenology of the bimodal model as it
pertains to astrophysical neutrino flux ratios and neutrinoless
double beta decay, respectively. The summary and conclusions are
given in Sec.~\ref{sect:conclusion}.

\section{Models with one or two Dirac masses and one-loop pseudo-Dirac-ness} \label{sect:models}

There are various possible gauge models in which the bimodal
possibility  for neutrinos can emerge naturally. In
Ref.~\cite{adm}, a model in which only a single mass eigenstate
has a Dirac mass was considered. We briefly discuss the key
ingredients of this model and also outline a different model in
which there could be two mass eigenstates with Dirac masses. These
models illustrate the point that the bimodal scenario leading to a
large Dirac and Majorana admixture for flavor neutrino states can
be realized within gauge models.

\subsection{One Dirac mass eigenstate}

Following the model in Ref.~\cite{adm}, an $S_3$ symmetry is
introduced,  permuting the three families of $SU(2)$ lepton
doublets $(L_e, L_\mu, L_\tau)$ among themselves. This reducible
representation of $S_3$ can be decomposed as
$\ul{3}=\ul{1}+\ul{2}$ so that the following linear combinations
of lepton doublet fields, transforming as one and two dimensional
representations of $S_3$, turn out to be the mass eigenstates:
\begin{align}
 L_2 &~=~ \frac{1}{\sqrt{3}}(L_e + L_\mu + L_\tau) \sim \ul{1} \, , \nonumber \\
(L_1,L_3) &~=~ \left(\frac{1}{\sqrt{6}}(2L_e - L_\mu -
L_\tau),\frac{1}{\sqrt{2}} (L_\mu - L_\tau)\right) \sim \ul{2} \, .
\label{eq:s3assign}
\end{align}
The $S_3$ singlet field couples to the right-handed (rh) neutrino field
$N_\mu$ (assumed to be an $S_3$ singlet), which is isolated from
the other two rh neutrinos by additional quantum numbers. This
could either be a $Z_n$ symmetry or may even be the local $B-L$
itself. For example, in Ref.~\cite{pleitez}, the $B-L$ quantum
number is chosen such that $N_\mu$ has $B-L  = - 5$ and $N_{e,\tau}$
each have $B-L  = + 4$. The $B-L$ breaking Higgs can be chosen to
have quantum numbers such that only $N_{e,\tau}$ have large
Majorana masses (see Ref.~\cite{adm} for details). After
integrating out the seesaw right-handed neutrinos $N_e$ and
$N_\tau$, the effective lepton Yukawa coupling and dimension five
terms can be written as
\begin{align}\label{lagrangian}
{\cal L}_\nu &~=~ h L_2 H_u N_\mu+\frac{h^{2}_1} {M_{N_e}}(L_1
H_u)^2+\frac{h^{2}_3}{M_{N_\tau}}(L_3 H_u)^2 + {\rm H.c.},
\end{align}
with $h$, $h_1$ and $h_3$ dimensionless coupling constants, and
$H_u$ the up-type Higgs doublet. After electroweak symmetry
breaking, the neutrino sector has one Dirac neutrino corresponding
to the mass eigenstate $\nu_2$, two Majorana eigenstates $\nu_1$
and $\nu_3$, as well as tribimaximal mixing (TBM) \cite{tbm}.

 The Lagrangian in Eq.~\eqref{lagrangian}, together with the $S_3$ assignments in Eq.~\eqref{eq:s3assign} and the group multiplication rules allow one to construct the symmetric $4\times 4$ neutrino mass matrix in the flavor basis $(\nu_e, \nu_\mu,\nu_\tau, N_\mu)$, i.e.,
\begin{align}
 M_\nu &= \frac{m_2}{\sqrt{3}}\left(\begin{array}{ccc|c} 0 & 0 & 0 & 1 \\ \cdot & 0 & 0 & 1 \\ \cdot & \cdot & 0 & 1 \\ \hline \cdot & \cdot & \cdot & 0 \end{array}\right) + \frac{m_1}{6}\left(\begin{array}{ccc|c} 4 & -2 & -2 & 0 \\ \cdot & 1 & 1 & 0 \\ \cdot & \cdot & 1 & 0 \\ \hline \cdot & \cdot & \cdot & 0 \end{array}\right) + \frac{m_3}{2}\left(\begin{array}{ccc|c} 0 & 0 & 0 & 0 \\ \cdot & 1 & -1 & 0 \\ \cdot & \cdot & 1 & 0 \\ \hline \cdot & \cdot & \cdot & 0 \end{array}\right) \nonumber \\[3mm]
 &= \left(\begin{array}{ccc|c} \frac{2m_1}{3} & -\frac{m_1}{3} & -\frac{m_1}{3} &\frac{m_2}{\sqrt{3}}\\[1.5mm] \cdot & \frac{m_1}{6}+\frac{m_3}{2} & \frac{m_1}{6}-\frac{m_3}{2} & \frac{m_2}{\sqrt{3}}\\[1.5mm] \cdot & \cdot & \frac{m_1}{6}+\frac{m_3}{2} &
\frac{m_2}{\sqrt{3}}\\[1.5mm] \hline \cdot & \cdot & \cdot & 0\end{array}\right),
\label{eq:mnutbm}
\end{align} 
with $m_1 = h_1^2 v_u^2/M_{N_e}$, $m_3 = h_3^2 v_u^2/M_{N_\tau}$, and  $m_2 = h v_u$, where $v_u = \langle H_u \rangle$. In order for the Majorana and Dirac mass matrix elements to have comparable magnitudes, the Yukawa coupling $h$ must be of order $10^{-12}$. This can be motivated in supersymmetric versions of such models, where the rh sneutrino drives inflation, and a small Dirac coupling is required to give consistent predictions (see Ref.~\cite{adm} for details). This bimodal scenario is in contrast to the usual pseudo-Dirac \cite{andre} models, in which the Majorana masses are much smaller than the Dirac masses, or to the seesaw mechanism, in which case the Dirac masses are much smaller than the right-handed Majorana mass scale. The implication is that despite the large Majorana mass terms, all oscillation results remain unaffected, unlike the conventional pseudo-Dirac case. In fact, it is easy to show that neutrinos described by this mass matrix propagate in matter in the same way as those in the pure Majorana or pure Dirac case. The propagation equation contains the active part of $M^\dagger_\nu M_\nu$, which has the same form for all the scenarios we are contemplating.

The matrix diagonalizing Eq.~\eqref{eq:mnutbm} to diag$(m_1,m_2,m_3,-m_2)$ is given by
\begin{eqnarray}
V_\nu = \left(\begin{array}{cc} U_{\rm TBM} & 0_3^T\\ 0_3 &
1\end{array}\right) \left(\begin{array}{cc} 1 & 0_3 \\ 0_3^T &
R(\pi/4)\end{array}\right), \label{eq:vnu}
\end{eqnarray}
where $0_3 = (0,0,0)$ and $R(\pi/4)$ is the $3\times 3$ unitary rotation matrix
\begin{equation}
  R(\pi/4) = \begin{pmatrix} \cos \frac{\pi}{4} & 0 & -\sin \frac{\pi}{4}
   \\ 0 & 1 & 0 \\ \sin \frac{\pi}{4} & 0 & \cos \frac{\pi}{4} \end{pmatrix}.
\end{equation}
A loop-induced pseudo-Dirac mass for $\nu_2$ (see
Sec.~\ref{subsect:loop}) will lead to perturbations to the
matrix in Eq.~(\ref{eq:mnutbm}). In the simple case where the
$(4,4)$ entry of $M_\nu$ is perturbed to $\epsilon \, m_2$, the
mixing matrix $V_\nu$ is modified to
\begin{eqnarray}
V'_\nu = \left(\begin{array}{cc} U_{\rm TBM} & 0_3^T\\ 0_3 & 1\end{array}\right)\left(\begin{array}{cc} 1 & 0_3 \\ 0_3^T & R(\pi/4 + \epsilon /4)\end{array}\right).
\label{eq:vpertnu}
\end{eqnarray}

In addition, the full Pontecorvo-Maki-Nakagawa-Sakata (PMNS) matrix will include rotations from the
charged lepton sector. This can be described by writing down the
most general Yukawa superpotential as
\begin{align}
{\cal W}_{l, Y} &~=~ \frac{1}{M}h_e H_d(L_e \sigma_e e^c+L_\mu
\sigma_\mu \mu^c + L_\tau \sigma_\tau \tau^c) \nonumber \\[2mm] &~+~ \frac{1}{M}h_\mu H_d
(L_\mu \sigma_e e^c+L_\tau \sigma_\mu \mu^c+L_e \sigma_\tau
\tau^c)  \\[2mm] &~+~ \frac{1}{M}h_\tau H_d (L_\tau \sigma_e e^c+L_e \sigma_\mu
\mu^c+L_\mu \sigma_\tau \tau^c) + {\rm H.c.}\, , \nonumber
\end{align}
where $(\sigma_e,\sigma_\mu,\sigma_\tau)$ are gauge singlet
superfields, $h_\alpha$ ($\alpha=e,\mu,\tau$) are the Yukawa
couplings, and we have assumed three extra $Z_n$ symmetries that
``glue'' each charged lepton singlet $(e^c,\mu^c,\tau^c)$ to the
corresponding $\sigma_\alpha$ gauge singlet \cite{s3tbm}. The
implication is that even though the neutrino mass matrix is
diagonalized by the TBM matrix, there are small corrections from
the charged lepton sector via the matrix $U_\ell$, which diagonalizes
it. As a result, the final PMNS matrix
$U_{\rm PMNS}$ has a perturbed TBM form, if one assumes the
charged lepton contributions to be small. The complete
diagonalization matrix takes the same form as
Eq.~\eqref{eq:vpertnu}, with $U_{\rm TBM}$ replaced by
$U_\ell^\dagger U_{\rm TBM}$.

In this model only the $\nu_2$ mass eigenstate has a Dirac mass,
but this idea can easily be generalized in the sense that one or
both of the other two neutrino mass eigenstates ($\nu_1$ and/or
$\nu_3$) could be Dirac. The case in which two of the eigenstates
are Dirac is motivated in the next subsection.

\subsection{Model with two Dirac mass eigenstates}

A model in which two Dirac mass eigenstates and one Majorana mass eigenstate appear naturally is a minimal $B-L$ extension of the minimal supersymmetric standard model (MSSM), where $B-L$ is broken by the vacuum expectation value (VEV) of the right-handed sneutrino~\cite{sogee}. Apart from the fact that there are three right-handed neutrinos (required for anomaly cancellation) and the gauge interactions associated with $B-L$, the model is essentially the same as the MSSM, i.e., two Higgs doublets that have zero $B-L$. This is therefore a minimal extension of the MSSM with local $B-L$. Furthermore, by choosing the gauge group to be $SU(2)_L\times U(1)_{I_{3R}}\times U(1)_{B-L}$, the model preserves gauge coupling unification with $B-L$ breaking at the TeV scale, without any additional fields. Radiative corrections can allow the sneutrino fields to acquire a nonzero VEV, for certain ranges of parameters. As was noted in the first reference of Ref.~\cite{sogee}, this leads to a neutrino mass matrix of the form
 \begin{eqnarray}\label{mv}
 M_\nu~=~\left(\begin{array}{ccc} 0_{3 \times 3} & h_\nu v_u & 0_3^T \\ h^T_\nu v_u & 0_{3 \times 3} &
g_{BL} \langle\tilde{\nu}^c_\alpha \rangle\\
 0_3 & g_{BL} \langle\tilde{\nu}^c_\alpha \rangle^T & \mu \end{array}\right) ,
\end{eqnarray}
where the rows and columns correspond to
$(\nu_\alpha,\nu^c_\alpha, \tilde{V})$, $\alpha = 1,2, 3$,  and
$\tilde{V}$ is the superpartner of the linear combination of $B-L$
and $I_{3R}$ gauge boson, i.e.~$(g_R V_{3R} - g_{BL}
V_{BL})/\sqrt{g^2_R+g^2_{BL}}$. Here $g_R, g_{BL}$ are the gauge
couplings of $U(1)_{I_{3R}}, U(1)_{B-L}$, respectively, $h_\nu$
is the $3\times 3$ Yukawa coupling matrix for $\nu^c$, and
$\langle\tilde{\nu}^c_\alpha \rangle$ is a column vector with
components given by the three $\tilde{\nu}^c$ VEVs. The
parameter
$\mu$ is the supersymmetry (SUSY) breaking Majorana mass term of the $B-L$ gaugino
$\tilde{V}$.
 One could in fact redefine the right-handed neutrino and sneutrino states so
that the linear combination of right-handed neutrinos that mixes with
left-handed neutrinos in the Dirac mass
in Eq.~(\ref{mv}) is same as the one for sneutrinos that picks up a VEV
(we can call this $\tilde{\nu}^c_e$). In that case, only one
right-handed neutrino is
kept in the $3\times 3$ neutrino mass matrix in the equation above.
The other two right-handed neutrinos remain coupled to the
left-handed neutrinos only through the Yukawa couplings and are not
present in Eq.~(\ref{mv}). In the
above mass matrix we have neglected small contributions that could
arise from induced sneutrino VEVs, which can mix the right-handed
neutrinos with Higgsinos.

It is clear that this matrix leads to one linear combination of
light neutrinos with Majorana mass given by the inverse seesaw
formula \cite{valle}, while the two other combinations only get a
Dirac mass. With additional symmetries, e.g.~$S_3$ as in
Ref.~\cite{adm}, one could get the TBM pattern for light
neutrinos. However, the main point for our discussion is that this
model can naturally lead to one Majorana and two Dirac mass
eigenstates.

For illustration, the full symmetric $5\times 5$ mass matrix in the flavor basis (for the TBM version where
 $\nu_1, \nu_2$ are Dirac type) would read
 \begin{eqnarray}
 M_\nu~=~\left(\begin{array}{ccc|cc} 0 & 0 & 0 &\frac{m_2}{\sqrt{3}} & \frac{2 m_1}{\sqrt{6}}\\[1.5mm]
 \cdot & \frac{m_3}{2} & -\frac{m_3}{2} & \frac{m_2}{\sqrt{3}} & -\frac{m_1}{\sqrt{6}}\\[1.5mm]
 \cdot & \cdot & \frac{m_3}{2} & \frac{m_2}{\sqrt{3}} & -\frac{m_1}{\sqrt{6}}\\[1.5mm] \hline \cdot & \cdot & \cdot
 & 0 & 0 \\[1.5mm] \cdot & \cdot & \cdot & \cdot & 0\end{array}\right) .
 \end{eqnarray}
In analogy to the case treated in the previous subsection, this
matrix can have Majorana mass terms that are similar in magnitude
to the Dirac masses.

The considerations from this and the preceding subsection indicate
that situations in which one or two neutrino mass eigenstates are
Dirac and the others (or other) Majorana are possible and arise in
simple models. We are therefore motivated to look for experimental
implications of these scenarios. However, we wish to note that the small Dirac masses are obtained at
the price of tuning the Yukawa couplings. Clearly it will be more
desirable to have a theory that can predict these small values.

\subsection{One-loop corrections to Dirac mass eigenstates} \label{subsect:loop}

Here we remark on the observation \cite{adm,petcov_1} that
one-loop corrections  to tree level Dirac state(s) pick up tiny
Majorana corrections, if lepton number is not conserved. For
simplicity, let us consider the case with one Dirac state; the
discussion easily generalizes to the case of two Dirac states.
Note that in the effective low energy Lagrangian of
Eq.~\eqref{lagrangian}, only a specific linear combination of the
lepton doublets that are eigenstates of $S_3$ appear. The charged
lepton mass terms break this symmetry, resulting in mixings
between different mass eigenstates in the finite wave function
renormalization corrections that arise at the one-loop level. In
the specific case of our first example, this will mean new terms
of the form $\delta_{12}\bar{\nu}_2\gamma^\mu\partial_\mu\nu_1$,
where $\delta_{12}\sim \frac{{G_F}m^2_\tau}{16\pi^2\sqrt{6}}\sim
10^{-7}$. Upon diagonalization of the kinetic terms, the new
states become $\nu'_1\approx \nu_1+\delta_{12}\nu_2$; hence the
Majorana mass term for $\nu_1$ in the new basis leads to a
Majorana mass of magnitude $\delta^2_{12}m_1$ for $\nu_2$. The
leading pseudo-Dirac contribution to the Dirac eigenstate is
therefore of order $10^{-14}\sqrt{\dma} \sim 10^{-15}$ eV
($\sqrt{\dma}$ is the atmospheric mass-squared difference),
corresponding to an oscillation length of $\sim 10$ kilo parsecs
(kpc). This implies that extra-galactic neutrinos from sources
beyond 10 kpc will have half of their $\nu_2$ component oscillate
into sterile neutrinos, thus affecting the observed flavor ratios
of extra-galactic neutrinos, which we discuss in the next section.

We will no longer specify the magnitude of the mass-squared
difference of the pseudo-Dirac neutrino $\nu_i$, but simply call
it $\delta m_i^2$, and analyze the phenomenological consequences
in flavor ratios at neutrino telescopes and in neutrinoless
double beta decay.

\section{Extra-galactic neutrino phenomenology} \label{sect:cosmonu}

 Extra-galactic neutrinos by definition travel large distances in
space and can have different energies, depending on their source.
Thus these neutrinos can be a probe of standard and nonstandard
neutrino properties (see Ref.~\cite{Pakvasa_reviews} for reviews).
In most cases \cite{Learned:1994wg} the neutrinos originate from
pion (and kaon) decay, followed by muon decay ($\pi^-\to \mu^- +
\bar{\nu}_\mu$ and $\mu^-\to e^-+\bar{\nu}_e+\nu_\mu$), giving the
initial flavor flux ratios of $\Phi^0_e:\Phi^0_\mu:\Phi^0_\tau =
1:2:0$. However, in neutron sources the initial ratios are
$1:0:0$, with electron antineutrinos originating from $\beta$
decays \cite{neutron_sources}; in muon-damped sources they become
$0:1:0$, since the muons (but not pions) lose energy before they
decay \cite{muon_sources}. Although the latter two sources are
presumably less common and harder to measure (there is less total
neutrino flux), they allow for interesting comparative studies
with the usual pure pion source.

In general, the initial flux composition may be described as
\cite{Choubey:2009jq} \be (\Phi_e^0 : \Phi_\mu^0 : \Phi_\tau^0) =
(1:n:0)\,. \label{eq:nparam} \ee Here the parameter $n$
distinguishes the different types of neutrino  sources:  for
neutron sources, the initial ratio of $1:0:0$ is represented by
the limit $n \rightarrow 0$, whereas in muon-damped sources the
initial ratio of $0:1:0$ is the limit $n \rightarrow \infty$. Pure
pion sources have $n=2$. In each case neutrino mixing will affect
the final flavor flux ratios at Earth detectors, and these ratios
will also depend on whether the neutrinos are pseudo-Dirac or
bimodal. It is well known that for the initial ratios of $1:2:0$,
the final ratios turn out to be $1:1:1$ \cite{Learned:1994wg},
assuming $\mu-\tau$ symmetry (actually, it suffices to assume that
$\Re\mathfrak{e}(U_{e3})=0$ and $\theta_{23}=\pi/4$) and three
standard neutrinos. Deviations from this symmetry limit will be
discussed below. If some or all of the neutrinos are pseudo-Dirac,
the detected flux ratios are modified, and it is possible to study
the effects of deviations in each different case.

In the standard three-neutrino scenario (no pseudo-Dirac effects), the flavor conversion probability reads
\begin{equation}
 P_{\alpha\beta} \equiv P(\nu_\alpha \rightarrow \nu_\beta) = \delta_{\alpha\beta}
- 2\sum_{i > j}\Re\mathfrak{e}\left(U_{\alpha j}U^*_{\alpha i} U^*_{\beta j} U_{\beta i}\right) =
\sum_i |U_{\alpha i}|^2|U_{\beta i}|^2 \, .
\label{eq:prob_usual}
\end{equation}
However, it can be shown that if all neutrinos are pseudo-Dirac \cite{Beacom:2003eu},
\begin{equation}
 P_{\alpha\beta} = \sum_i |U_{\alpha i}|^2|U_{\beta i}|^2
\cos^2\left(\frac{\delta m_i^2 L}{4E}\right),
\label{eq:prob_pD}
\end{equation}
where $\delta m_i^2=(m_i^+)^2-(m_i^-)^2$ is the small mass-squared
difference  between the pseudo-Dirac pairs, and it is obvious that
this reduces to Eq.~\eqref{eq:prob_usual} for $\delta m_i^2=0$. In
the spirit of the bimodal flavor neutrino cases discussed above,
not all of the three states $\nu_i$ could be pseudo-Dirac, but only
one or two. If the corresponding $\delta m_i^2 L/4E \gg 1$, in
other words if $L/E$ is large enough, the cosine term averages out
to $1/2$. The standard effects from neutrino mixing are therefore
modified and neutrinos from very distant sources could probe the
tiny pseudo-Dirac mass-squared differences
\cite{Beacom:2003eu,Keranen:2003xd,Esmaili:2009fk,Bhattacharya:2010xj}.
Recall that mass splittings of less than about $10^{-11}$~eV$^2$
have no effect on the solar neutrino flux \cite{andre}.

If one assumes that only one neutrino is pseudo-Dirac (say $\nu_2$), then the
corresponding term ($i=2$) of the sum in Eq.~\eqref{eq:prob_pD} is modified by a
factor of $1/2$, leading to the probability
\begin{equation}
 P_{\alpha\beta} = |U_{\alpha 1}|^2|U_{\beta 1}|^2+\frac{1}{2}|U_{\alpha 2}|^2|U_{\beta 2}|^2
+|U_{\alpha 3}|^2|U_{\beta 3}|^2\ .
\label{eq:prob_pDnu2}
\end{equation}
This can be extended to cases in which different combinations of neutrinos are pseudo-Dirac;
the reduction factor of $1/2$ is applied to the relevant terms in each case. The measured
neutrino flux, $\Phi_\alpha$, is the sum of the product of each initial flux $\Phi_\alpha^0$
with the relevant flavor conversion probability,
\begin{equation}
 \Phi_\alpha = \sum_\beta P_{\beta\alpha}\Phi_\beta^0\ ,
\end{equation}
so that the presence of one or more pseudo-Dirac neutrinos will
change the final detected flux (and flux ratios) compared to the
standard case. Table~\ref{table:mutauratios} shows the observable
$\ratiomue$ ratio as a function of $\theta_{12}$ for the different
combinations of pseudo-Dirac neutrinos, for $\mu-\tau$ symmetry
and initial fluxes of $1:2:0$. Note that if all three neutrinos
are pseudo-Dirac the observed flux ratio is again $1:1$, with an
overall reduction in flux of $1/2$. In several cases the ratio is
independent  of $\theta_{12}$. We also observe that if $\nu_2$ or
$\nu_{1,3}$ are pseudo-Dirac, then $\ratiomue$ is $1:1$ only if
$\sin^2 \theta_{12} = \frac 13$, i.e., for exact TBM.
\begin{table}[b]
\centering
 \caption{The observed $\ratiomue$ neutrino flux ratio at the detector for
different combinations of pseudo-Dirac neutrinos, assuming the
initial flux ratios of $1:2:0$ and exact $\mu-\tau$ symmetry.
Numerical values are calculated from the global fit data in
Ref.~\cite{globaloscdata}.}
 \label{table:mutauratios}
\vspace{10pt}
 \begin{tabular}{lccc}
 \hline \hline
\T \multirow{2}{*}{Pseudo-Dirac neutrinos} & \multicolumn{3}{c}{$\ratiomue$} \\
 & General case & Best-fit & $3\sigma$\\[2mm]
\hline \T None \& all & $1:1$ & 1.00 & 1.00 \\[2mm]
$\nu_1$ & $1-\frac{1}{4}\sss:\frac{1}{2}(1+\sss)$ & 1.40 & $1.32\!-\!1.46$ \\[2mm]
$\nu_2$ \& $\nu_3$ & $\frac{1}{4}(2+\sss):1-\frac{1}{2}\sss$ & 0.67 & $0.66\!-\!0.73$ \\[2mm]
$\nu_2$ & $\frac{1}{4}(3+\sss):1-\frac{1}{2}\sss$ & 0.99 & $0.95\!-\!1.04$ \\[2mm]
$\nu_1$ \& $\nu_3$ & $\frac{1}{4}(3-\sss):\frac{1}{2}(1+\sss)$ & 0.58 & $0.55\!-\!0.60$ \\[2mm]
$\nu_3$ & $\frac{3}{4}:1$ & 0.75 & 0.75 \\[2mm]
$\nu_1$ \& $\nu_2$ & $\frac{3}{2}:1$ & 1.50 & 1.50\\[2mm]
\hline \hline
\end{tabular}
\end{table}

In the most general case one can expect deviations from exact
$\mu-\tau$ symmetry, so
that the relations $\theta_{13}=0$ and $\theta_{23}=\frac{\pi}{4}$ are not exact.
Defining the deviation parameter
\begin{equation}
 \epsilon = \frac{\pi}{4}-\theta_{23}\, ,
\end{equation}
the probability matrix $P$ (without pseudo-Dirac effects) with elements
$P_{\alpha \beta}$ can be approximated as
\begin{equation}
 P \approx \begin{pmatrix} 1-2c_{12}^2s_{12}^2 & c_{12}^2 s_{12}^2 +
  \Delta & c_{12}^2s_{12}^2 - \Delta \\ \cdot &
  \frac{1}{2}(1-c_{12}^2s_{12}^2)-\Delta & \frac{1}{2}(1-c_{12}^2s_{12}^2) \\
   \cdot & \cdot & \frac{1}{2}(1-c_{12}^2s_{12}^2) + \Delta \end{pmatrix}, \label{eq:genprobmatrix}
\end{equation}
where $s_{12}^2 = \sss$, $c_{12}^2 = \css$ and the universal
correction parameter is defined as
\cite{Xing:2006xd,Rodejohann:2006qq}
\begin{equation}
 \Delta \equiv \frac{1}{4}(2\epsilon \sts + \theta_{13}\cos\delta\sin 4\theta_{12})
  = 2\epsilon s_{12}^2 c_{12}^2 + \frac{1}{4}\theta_{13}\cos\delta\sin 4\theta_{12}\,.
\end{equation}
Terms of order $\mathcal{O}(\theta_{13}^2)$,
$\mathcal{O}(\epsilon^2)$, and $\mathcal{O}(\theta_{13}\epsilon)$
have been neglected in this approximation. In this case it can be
shown that the  flux ratio evolves as
\cite{Xing:2006xd,Rodejohann:2006qq}
\begin{equation}
 (1:2:0) \longrightarrow (1+2\Delta):(1-\Delta):(1-\Delta) \, .
\label{eq:genfluxratiosneq2}
\end{equation}
The parameters $\epsilon$ and $\Delta$ lie in the ranges
$-0.148~\leq\epsilon\leq0.166$ and $-0.10~\leq~\Delta~\leq 0.11$
for the current $3\sigma$ ranges \cite{globaloscdata} of the
oscillation parameters.\footnote{``Next-to-next-to-leading order'' terms of second order in
$\epsilon$ and $\theta_{13} $ have been discussed in
\cite{Pakvasa:2007dc,Meloni:2006gv}.}

This general framework can be applied to the pseudo-Dirac
scenario:  if one or more neutrinos are pseudo-Dirac, the
probability matrix in Eq.~\eqref{eq:genprobmatrix} will be
modified, leading to different final flux ratios in each case.
These probabilities can be written (to first order in
$\theta_{13}$ and $\epsilon$) in terms of $\theta_{12}$,
$\theta_{13}$, and the deviation parameters $\epsilon$, $\Delta$,
and $\Gamma$, where
\begin{equation}
 \Gamma \equiv \frac{1}{8}\theta_{13}\sin2\theta_{12} \cos\delta = \frac{1}{4}\theta_{13} s_{12} c_{12}\cos\delta\, .
\end{equation}
The parameter $\Gamma$ is constrained to the range  $-0.026 \leq
\Gamma \leq 0.026$. For each case, the flavor conversion
probabilities are given by\footnote{The ratio $P_{\tau\tau}$ is
omitted, as it is not needed to calculate flux ratios.}
\begin{itemize}
 \item \textbf{$\nu_1$ pseudo-Dirac:}
\begin{align}
 P^{\nu_1}_{ee} &= \frac{1}{2}c^4_{12}+s^4_{12}\, , \nonumber \\
 P^{\nu_1}_{e\mu} &= \frac{3}{4}(c^2_{12}s^2_{12} + \Delta) - \Gamma\, , \nonumber \\
 P^{\nu_1}_{\mu\mu} &= \frac{1}{2}-\frac{1}{8}s_{12}^2\left(1+3c_{12}^2+4\epsilon\right) - \frac{3}{4}\Delta - \Gamma\, ,  \\
 P^{\nu_1}_{e\tau} &= \frac{3}{4}(c^2{12}s^2_{12} -\Delta) + \Gamma\, , \nonumber \\
 P^{\nu_1}_{\mu\tau} &= \frac{1}{4} + \frac{1}{4}c^4_{12} + \frac{1}{8}s^4_{12}\, , \nonumber
\end{align}
 \item \textbf{$\nu_2$ and $\nu_3$ pseudo-Dirac:}
\begin{align}
 P^{\nu_{2,3}}_{ee} &= c^4_{12} + \frac{1}{2}s^4_{12}\, , \nonumber \\
 P^{\nu_{2,3}}_{e\mu} &= \frac{3}{4}(c^2_{12}s^2_{12} + \Delta) + \Gamma\, , \nonumber \\
 P^{\nu_{2,3}}_{\mu\mu} &= \frac{1}{4}+\frac{1}{8}s_{12}^2(1-3c_{12}^2+4\epsilon) - \frac{3}{4}\Delta + \Gamma\, ,  \\
 P^{\nu_{2,3}}_{e\tau} &= \frac{3}{4}(c^2_{12}s^2_{12} -\Delta) - \Gamma\, , \nonumber \\
 P^{\nu_{2,3}}_{\mu\tau} &= \frac{1}{8} + \frac{1}{8}c^4_{12} + \frac{1}{4}s^4_{12}\, , \nonumber
\end{align}
 \item \textbf{$\nu_2$ pseudo-Dirac:}
\begin{align}
 P^{\nu_{2}}_{ee} &= c^4_{12} + \frac{1}{2}s^4_{12}\, , \nonumber \\
 P^{\nu_{2}}_{e\mu} &= \frac{3}{4}(c^2_{12}s^2_{12} + \Delta) + \Gamma\, , \nonumber \\
 P^{\nu_{2}}_{\mu\mu} &= \frac{1}{2}-\frac{1}{8}c_{12}^2(1+3s_{12}^2+4\epsilon) - \frac{3}{4}\Delta + \Gamma\, ,  \\
 P^{\nu_{2}}_{e\tau} &= \frac{3}{4}(c^2_{12}s^2_{12} - \Delta) - \Gamma\, , \nonumber \\
 P^{\nu_2}_{\mu\tau} &= \frac{1}{4} + \frac{1}{8}c^4_{12} + \frac{1}{4}s^4_{12}\, , \nonumber
\end{align}
 \item \textbf{$\nu_1$ and $\nu_3$ pseudo-Dirac:}
\begin{align}
 P^{\nu_{1,3}}_{ee} &= \frac{1}{2}c^4_{12}+s^4_{12}\, , \nonumber \\
 P^{\nu_{1,3}}_{e\mu} &= \frac{3}{4}(c^2_{12}s^2_{12} + \Delta) - \Gamma\, , \nonumber \\
 P^{\nu_{1,3}}_{\mu\mu} &= \frac{1}{4}+\frac{1}{8}c_{12}^2(1-3s_{12}^2+4\epsilon) - \frac{3}{4}\Delta - \Gamma\, , \\
 P^{\nu_{1,3}}_{e\tau} &= \frac{3}{4}(c^2_{12}s^2_{12} - \Delta) + \Gamma\, , \nonumber \\
 P^{\nu_{1,3}}_{\mu\tau} &= \frac{1}{8} + \frac{1}{4}c^4_{12} + \frac{1}{8}s^4_{12}\, , \nonumber
\end{align}
 \item \textbf{$\nu_3$ pseudo-Dirac:}
\begin{align}
 P^{\nu_3}_{ee} &= c^4_{12}+s^4_{12}\, , \nonumber \\
 P^{\nu_3}_{e\mu} &= c_{12}^2s_{12}^2 + \Delta\, , \nonumber \\
 P^{\nu_3}_{\mu\mu} &= \frac{3}{8} + \frac{1}{2}\left(c_{12}^2s_{12}^2+\epsilon\right)-\Delta\, , \\
 P^{\nu_3}_{e\tau} &= c_{12}^2s_{12}^2 -\Delta\, ,\nonumber \\
 P^{\nu_3}_{\mu\tau} &= \frac{3}{8} -\frac{1}{2}c_{12}^2s^2_{12}\, , \nonumber
\end{align}
 \item \textbf{$\nu_1$ and $\nu_2$ pseudo-Dirac:}
\begin{align}
 P^{\nu_{1,2}}_{ee} &= \frac{1}{2}(c^4_{12}+s^4_{12})\, , \nonumber \\
 P^{\nu_{1,2}}_{e\mu} &= \frac{1}{2}(c_{12}^2s_{12}^2 + \Delta)\, , \nonumber \\
 P^{\nu_{1,2}}_{\mu\mu} &= \frac{3}{8} - \frac{1}{4}c_{12}^2s_{12}^2 -\frac{1}{2}\epsilon -\frac{1}{2}\Delta\, , \\
 P^{\nu_{1,2}}_{e\tau} &= \frac{1}{2}(c_{12}^2s_{12}^2 -\Delta)\, , \nonumber \\
 P^{\nu_{1,2}}_{\mu\tau} &= \frac{3}{8} -\frac{1}{4}c_{12}^2s^2_{12}\, . \nonumber
\end{align}
\end{itemize}
These expressions can be used to calculate the final flux ratios
in each case. We discuss  the most straightforwardly measurable
flux ratio $\ratiomue$ and also display results for the ratio
$\ratioetau$, which is harder to measure.

The plots in Figs.~\ref{fig:ratio_1e2mu} and \ref{fig:tauratio_1e2mu} show the variation  in the flux ratios
$\ratiomue$ and $\ratioetau$ with $\sss$ for the different possible combinations of one or two pseudo-Dirac neutrinos,
assuming the standard case of an initial flux ratio of $1:2:0$ (a
pure pion source). For comparison, the standard case without any
pseudo-Dirac nature is also shown, for which the ratios can be
approximated by
\begin{equation}
 \frac{\Phi_\mu}{\Phi_e} \approx \frac{1-\Delta}{1+2\Delta} \approx 1-3\Delta \quad {\rm and} \quad \frac{\Phi_e}{\Phi_\tau} \approx \frac{1+2\Delta}{1-\Delta} \approx 1+3\Delta\, ,
\label{eq:fluxrationeq2}
\end{equation}
using Eq.~\eqref{eq:genfluxratiosneq2} and neglecting quadratic
terms. One can see from the plots in
Figs.~\ref{fig:ratio_1e2mu} and \ref{fig:tauratio_1e2mu} that the two cases
in which $\nu_3$ and $\nu_{1,2}$ are pseudo-Dirac show very little dependence
 on $\theta_{12}$ (compare with Table~\ref{table:mutauratios}), even with deviations
 applied. 
The ratio $\ratiomue$ differs considerably from
the standard case if either $\nu_1$ or both $\nu_1$ and $\nu_2$
are pseudo-Dirac, and can be approximated by
\begin{align}
 \frac{\Phi^{\nu_1}_\mu}{\Phi^{\nu_1}_e} &\approx \{1 - 3\Delta\} + \frac{2-s_{12}^2-3s_{12}^4}{2(1+s_{12}^2)^2} -\frac{2\epsilon(s_{12}^2+s_{12}^4)}{(1+s_{12}^2)^2} -\frac{3\Delta(3-4s_{12}^2-2s_{12}^4)}{2(1+s_{12}^2)^2}+\frac{2\Gamma(1-4s_{12}^2)}{(1+s_{12}^2)^2}\, , \\[2mm]
  \frac{\Phi^{\nu_{1,2}}_\mu}{\Phi^{\nu_{1,2}}_e} &\approx \{1 - 3\Delta\} +\frac{1}{2} - 2\epsilon -\Delta\, .
\end{align}
In both cases the expressions are given to first order in the
deviation  parameters and  the curly brackets correspond to the
standard case [Eq.~\eqref{eq:fluxrationeq2}]. For the ratio
$\ratioetau$, Fig.~\ref{fig:tauratio_1e2mu} shows that there are
potentially strong effects if either $\nu_3$ or both $\nu_2$ and
$\nu_3$  are pseudo-Dirac, in which case
\begin{align}
 \frac{\Phi^{\nu_3}_e}{\Phi^{\nu_3}_\tau} &\approx \{1 + 3\Delta\} + \frac{1}{3} + \frac{13}{9}\Delta\, , \\[2mm]
 \frac{\Phi^{\nu_{2,3}}_e}{\Phi^{\nu_{2,3}}_\tau} &\approx \{1 + 3\Delta\} + \frac{8 + 4s_{12}^2-8c_{12}^2s_{12}^2-18s_{12}^4-3s_{12}^6}{(2+s_{12}^2)^3} + \frac{3\Delta(8+4s_{12}^2-6s_{12}^4-s_{12}^6)}{(2+s_{12}^2)^3} + \frac{32\Gamma(2+s_{12}^2)}{(2+s_{12}^2)^3}\ .
\end{align}

 As mentioned above, the initial flavor ratios of other
interesting neutrino sources, such as neutron or muon-damped
sources, can be parameterised as in Eq.~\eqref{eq:nparam}.
Figures~\ref{fig:ratio_1enmu} and \ref{fig:tauratio_1enmu}
indicate the dependence of the ratios $\ratiomue$ and $\ratioetau$
on $n$ for the different pseudo-Dirac combinations. It is evident
that in certain cases the observed ratio can be much larger than
in the standard case. Specifically, in the case of $\nu_2$ being
pseudo-Dirac, the ratios $\ratiomue$ and $\ratioetau$ can become
large for $n \rightarrow \infty$ and $n \rightarrow 0$,
respectively. Expanding to first order in the deviation
parameters, the ratios are given in these cases by
\begin{align}
 \frac{\Phi^{\nu_2}_\mu}{\Phi^{\nu_2}_e} &\xrightarrow{n\rightarrow\infty} \frac{P^{\nu_2}_{\mu\mu}}{P^{\nu_2}_{e\mu}}
 \approx \left\{\frac{1-c_{12}^2s_{12}^2}{2c_{12}^2s_{12}^2} - \frac{1+c_{12}^2s_{12}^2}{2c_{12}^4s_{12}^4}\Delta\right\} -\frac{1+4\epsilon}{6s_{12}^2} +\frac{1+12\Gamma}{6c_{12}^2s_{12}^2} + \frac{3\Delta+4\Gamma}{18c_{12}^2s_{12}^4} - \frac{3\Delta+16\Gamma}{18c_{12}^4s_{12}^4} \, , \\[2mm]
 \frac{\Phi^{\nu_2}_e}{\Phi^{\nu_2}_\tau} &\xrightarrow{n\rightarrow 0} \frac{P^{\nu_2}_{ee}}{P^{\nu_2}_{e\tau}}
 \approx \left\{\frac{1-2c_{12}^2s_{12}^2}{c_{12}^2s_{12}^2} + \frac{1-2c_{12}^2s_{12}^2}{c_{12}^4s_{12}^4}\Delta\right\} + \frac{1}{3c_{12}^2s_{12}^2} -\frac{6\Delta+8\Gamma}{9c_{12}^4s_{12}^2} + \frac{3\Delta+16\Gamma}{9c_{12}^4s_{12}^4}\ .
\end{align}
In each case the terms in curly brackets again denote the flux
ratios corresponding to the general case, without pseudo-Dirac
neutrinos, in the same  limit ($n\rightarrow\infty$ or
$n\rightarrow 0$). Additionally, if both $\nu_1$ and $\nu_2$ are
pseudo-Dirac, the ratio $\ratiomue$ becomes large for $n
\rightarrow \infty$,
\begin{align}
 \frac{\Phi^{\nu_{1,2}}_\mu}{\Phi^{\nu_{1,2}}_e} \xrightarrow{n\rightarrow\infty} \frac{P^{\nu_{1,2}}_{\mu\mu}}{P^{\nu_{1,2}}_{e\mu}} \approx \left\{\frac{1-c_{12}^2s_{12}^2}{2c_{12}^2s_{12}^2} - \frac{1+c_{12}^2s_{12}^2}{2c_{12}^4s_{12}^4}\Delta\right\} +\frac{1-4\epsilon}{4c_{12}^2s_{12}^2}-\frac{1}{c_{12}^4s_{12}^4}\Delta\ ,
\end{align}
and if both $\nu_2$ and $\nu_3$ are pseudo-Dirac, the ratio $\ratioetau$ becomes large for $n \rightarrow 0$,
\begin{equation}
 \frac{\Phi^{\nu_{2,3}}_e}{\Phi^{\nu_{2,3}}_\tau} \xrightarrow{n\rightarrow 0} \frac{P^{\nu_{2,3}}_{ee}}{P^{\nu_{2,3}}_{e\tau}} = \frac{P^{\nu_2}_{ee}}{P^{\nu_2}_{e\tau}} \approx \left\{\frac{1-2c_{12}^2s_{12}^2}{c_{12}^2s_{12}^2} + \frac{1-2c_{12}^2s_{12}^2}{c_{12}^4s_{12}^4}\Delta\right\} + \frac{1}{3c_{12}^2s_{12}^2} -\frac{6\Delta+8\Gamma}{9c_{12}^4s_{12}^2} + \frac{3\Delta+16\Gamma}{9c_{12}^4s_{12}^4}\ .
\end{equation}
The plots in Figs.~\ref{fig:ratio_1enmu} and
\ref{fig:tauratio_1enmu} could  in principle be used to rule out
certain cases. If, for instance, measurements of the neutrino flux
ratios from a muon-damped source give $\ratiomue \gtrsim 5$, four
of the six possibilities would be ruled out so that either
$\nu_2$ or both $\nu_1$ and $\nu_2$ would have to be pseudo-Dirac
neutrinos. A similar result applies for the case of $\ratioetau
\gtrsim 7$ and neutron sources, where only $\nu_2$ or $\nu_2$ and
$\nu_3$ could be pseudo-Dirac.

\section{neutrinoless double beta decay phenomenology} \label{sect:0nubb}

Another experimental test of the bimodal flavor neutrino
scenario is neutrinoless  double beta decay (see
Ref.~\cite{0nubb_reviews} for reviews). In the general case with
three Majorana neutrino mass eigenstates, the amplitude for this
process is proportional to the effective Majorana mass
\begin{align}
 \mee = \left|\sum_{i=1}^3 U_{ei}^2 m_i\right| \label{eq:meegen} 
 = \left| c_{12}^2c_{13}^2 |m_1| + s_{12}^2c_{13}^2 |m_2|e^{i\alpha} + s_{13}^2 |m_3|e^{i\beta}\right|,
\end{align}
with $\alpha$ and $\beta$ the Majorana phases. Here the decay is
mediated by light, active, and massive Majorana neutrinos, and one
assumes that there are no other new physics contributions, such as
heavy neutrino exchange, right-handed currents or the exchange of supersymmetric particles.

The coherent sum $\mee$ contains 7 out of 9 parameters of the
neutrino mass matrix and is the only observable carrying
information about the Majorana phases. It is possible for $\mee$
to vanish in the case of normal neutrino mass ordering; this is
equivalent to a zero in the (1,1) element of the low energy
Majorana neutrino mass matrix. In the inverted ordering case
$\mee$ cannot vanish, and the lower limit is given by
\begin{equation}
  \mee \approx \left|(c_{12}^2+s_{12}^2e^{i\alpha})\sqrt{\dma}\,\right| \gs
(c_{12}^2 - s_{12}^2 )\sqrt{\dma} \approx
\frac{\sqrt{\dma}}{3} \approx 17 \ {\rm meV} ,
\label{eq:meeminihstandard}
\end{equation}
where $\dma$ is the mass-squared difference of atmospheric
neutrinos.  The final extraction of the decay half-life is
affected by uncertainties in the nuclear matrix elements so that
an indisputable measurement of this process requires improved
precision in both particle and nuclear physics parameters. Future
experiments such as GERDA and SuperNEMO aim to reach a sensitivity
of order 10 meV and should thus be able to rule out the inverted
mass ordering, as long as the relevant parameter uncertainties are
reduced.

The standard picture of neutrinoless double beta decay is
modified in the presence of pseudo-Dirac neutrinos. With three
pseudo-Dirac neutrinos, the expression in Eq.~\eqref{eq:meegen}
becomes proportional to $\sum_{i=1}^3 U_{ei}^2 \frac{\delta
m_i^2}{2m_i}$
because the approximately degenerate eigenstates of the
pseudo-Dirac  pair have opposite CP parities. This contribution is
effectively vanishing ($\mee \lesssim 10^{-4}$ eV) and can be
neglected. However, if only one or two neutrino mass eigenstates
are pseudo-Dirac (the bimodal scenario), one effectively has a
combination of the standard case [Eq.~\eqref{eq:meegen}] and the
pure pseudo-Dirac case. Those neutrinos that are pseudo-Dirac do
not contribute to $\mee$, whereas the normal Majorana mass
eigenstates contribute as in Eq.~\eqref{eq:meegen}.

Figs.~\ref{fig:mee_sum_1dirac}~and~\ref{fig:mee_sum_1majorana}
show the allowed ranges in $\mee-\sumnu$ parameter space, for
different combinations of pseudo-Dirac neutrinos and both normal
and inverted neutrino mass ordering. The parameter space in
the standard case is included for comparison. We have plotted the
effective mass against the sum of masses $\sumnu$, rather than the
smallest mass itself, because the latter is, strictly speaking,
not an observable.

In each case, the contribution from the pseudo-Dirac pair was assumed to be vanishing so that
\begin{equation}
 \mee = \left|\sum_{j=1}^N U_{ej}^2 m_j\right| ,
\label{eq:meepdplots}
\end{equation}
where the index $j$ runs over the neutrinos that are {\em not}
pseudo-Dirac, and $N=1$ or $N=2$.
For instance, in the case where only $\nu_2$ is pseudo-Dirac, the effective Majorana mass becomes
\begin{equation}
 \mee = \left| c_{12}^2c_{13}^2 |m_1| + s_{13}^2 |m_3|e^{i\beta}\right|,
\end{equation}
and there is only one phase, $\beta$. One can see from the plots in
Figs.~\ref{fig:mee_sum_1dirac} and \ref{fig:mee_sum_1majorana} that in
the cases of $\nu_2$ and $\nu_{2,3}$ pseudo-Dirac and inverted mass
ordering, the lower limit for $\mee$ is increased by a factor of
2 \cite{adm}. Explicitly, 
the lower bound for the inverted ordering becomes
\begin{equation}
  \mee \approx c_{12}^2\sqrt{\dma} \gs \frac{2\sqrt{\dma}}{3} \approx
34 \ {\rm meV} \, ,
\end{equation}
to be compared with the bound for the standard case in Eq.~\eqref{eq:meeminihstandard}.

Because of the fact that $c_{12}^2 - s_{12}^2 \approx s_{12}^2$, the
case in which $\nu_1$ is pseudo-Dirac results in $\mee$~taking its
minimal value in the inverted ordering. Another interesting case
is when $\nu_{1,3}$ are pseudo-Dirac with normal ordering, where
the lower limit of $\mee$ is given by ($\dms$ is the mass-squared
difference of solar neutrinos)
\begin{equation}
  \mee \gs s_{12}^2\sqrt{\dms} \approx 2.9 \ {\rm meV}\, ,
\end{equation}
and the amplitude for double beta decay can never vanish, in contrast to the usual normal ordering case.

The cases where both $\nu_1$ and $\nu_2$ are pseudo-Dirac
obviously  lead to small values of $\mee$, since the only term
contributing is $s_{13}^2 |m_3|$. In these cases the effective
mass can lie outside the regions in which one expects it in the
general case. Another interpretation of this would be that one of
the nonstandard mechanisms of neutrinoless double beta decay
destructively interferes with the usual mass mechanism. The
strategy to test this would be to perform multi-isotope
investigation, as the cancellation is not expected to be on the
same level in different nuclei. However, the pseudo-Dirac
suppression discussed here is the same for all nuclei.

In summary, there are several cases 
for which there is a significant difference from the standard case of
pure Majorana neutrinos. If long baseline oscillation experiments
establish the neutrino ordering, and/or the neutrino mass scale is
pinned down by cosmology or direct searches, neutrinoless double
decay can distinguish the different  cases. This illustrates the
discriminative power of the process.

\section{Conclusion} \label{sect:conclusion}

\begin{figure}[htp]
 \centering
 \includegraphics[width=0.8\textwidth]{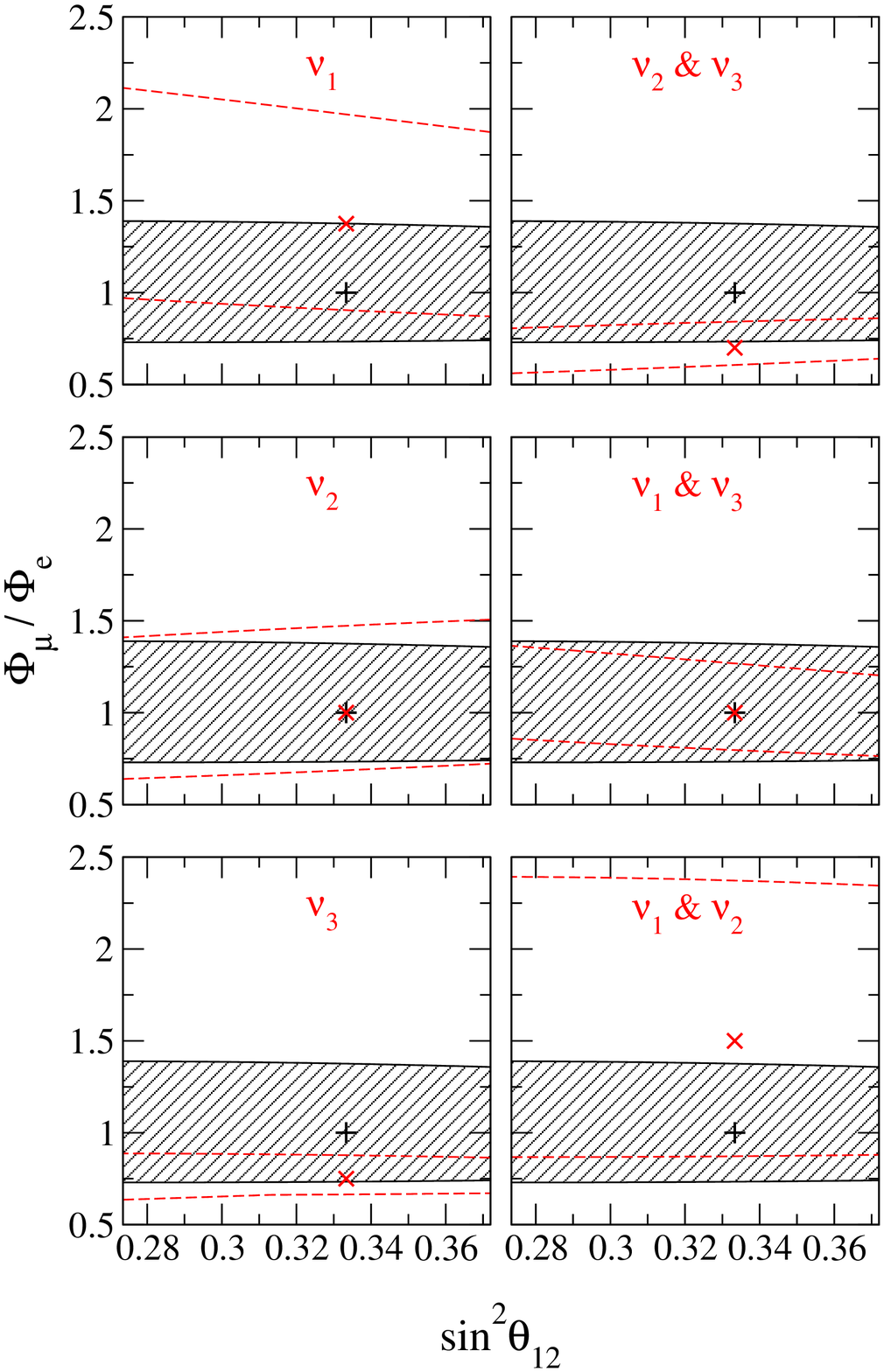}
 \caption{The observable flux ratio $\ratiomue$ against $\sss$, assuming the
initial neutrino flux ratios of $1:2:0$ and different combinations of
pseudo-Dirac neutrinos (denoted by red dashed lines), with the parameters
$\theta_{13}$, $\theta_{23}$, and $\delta$ varying in their allowed $3\sigma$ ranges.
The black hatched region shows the general case with no pseudo-Dirac neutrinos;
the red cross (black plus sign) shows the value of $\ratiomue$ in each
pseudo-Dirac case (the general case), assuming TBM.}
\label{fig:ratio_1e2mu}
\end{figure}

\begin{figure}[htp]
 \centering
 \includegraphics[width=0.8\textwidth]{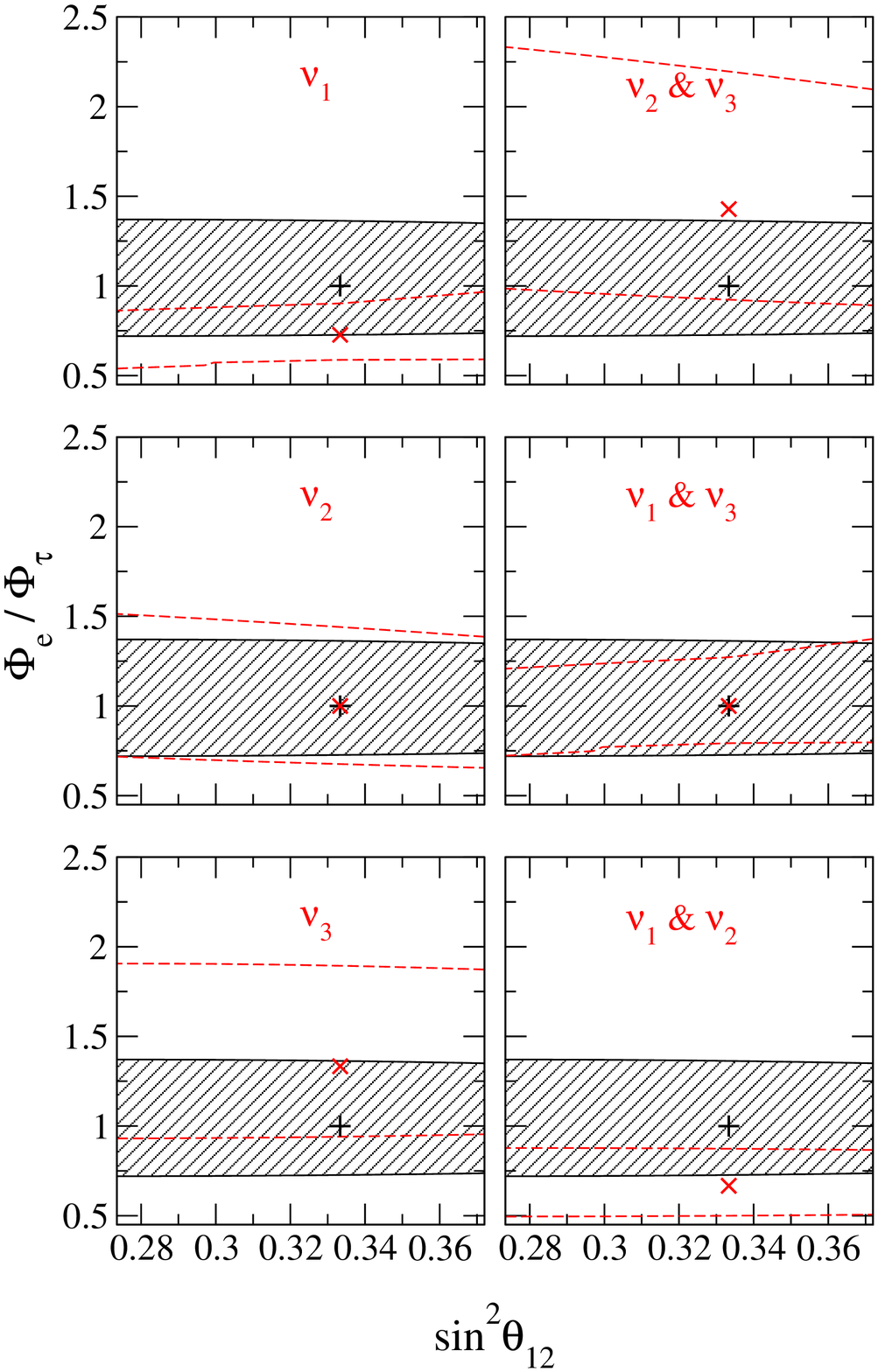}
 \caption{Same as Fig.~\ref{fig:ratio_1e2mu} for the observable flux ratio $\ratioetau$.}
\label{fig:tauratio_1e2mu}
\end{figure}

\begin{figure}[htp]
 \centering
 \includegraphics[width=0.8\textwidth]{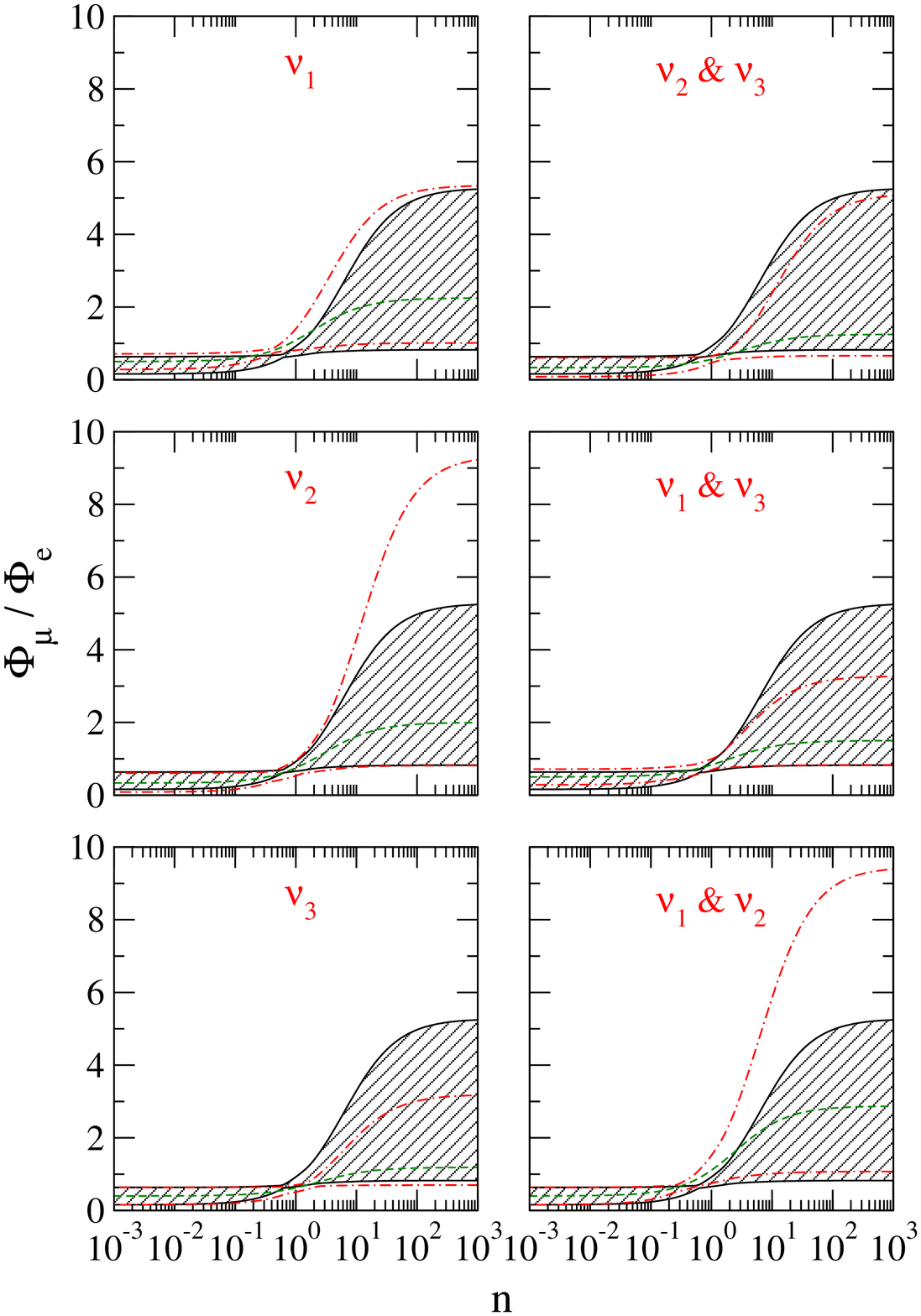}
 \caption{The observable flux ratio $\ratiomue$ against $n$, assuming the initial neutrino
flux ratios of $1:n:0$ and different combinations of pseudo-Dirac neutrinos
(denoted by red dashed-dotted lines), with the parameters $\theta_{13}$, $\theta_{23}$,
and $\delta$ varying in their allowed $3\sigma$ ranges. The black hatched region
shows the general case with no pseudo-Dirac neutrinos; the green dashed
line shows the value of $\ratiomue$ in each pseudo-Dirac case, assuming TBM.}
\label{fig:ratio_1enmu}
\end{figure}

\begin{figure}[htp]
 \centering
 \includegraphics[width=0.8\textwidth]{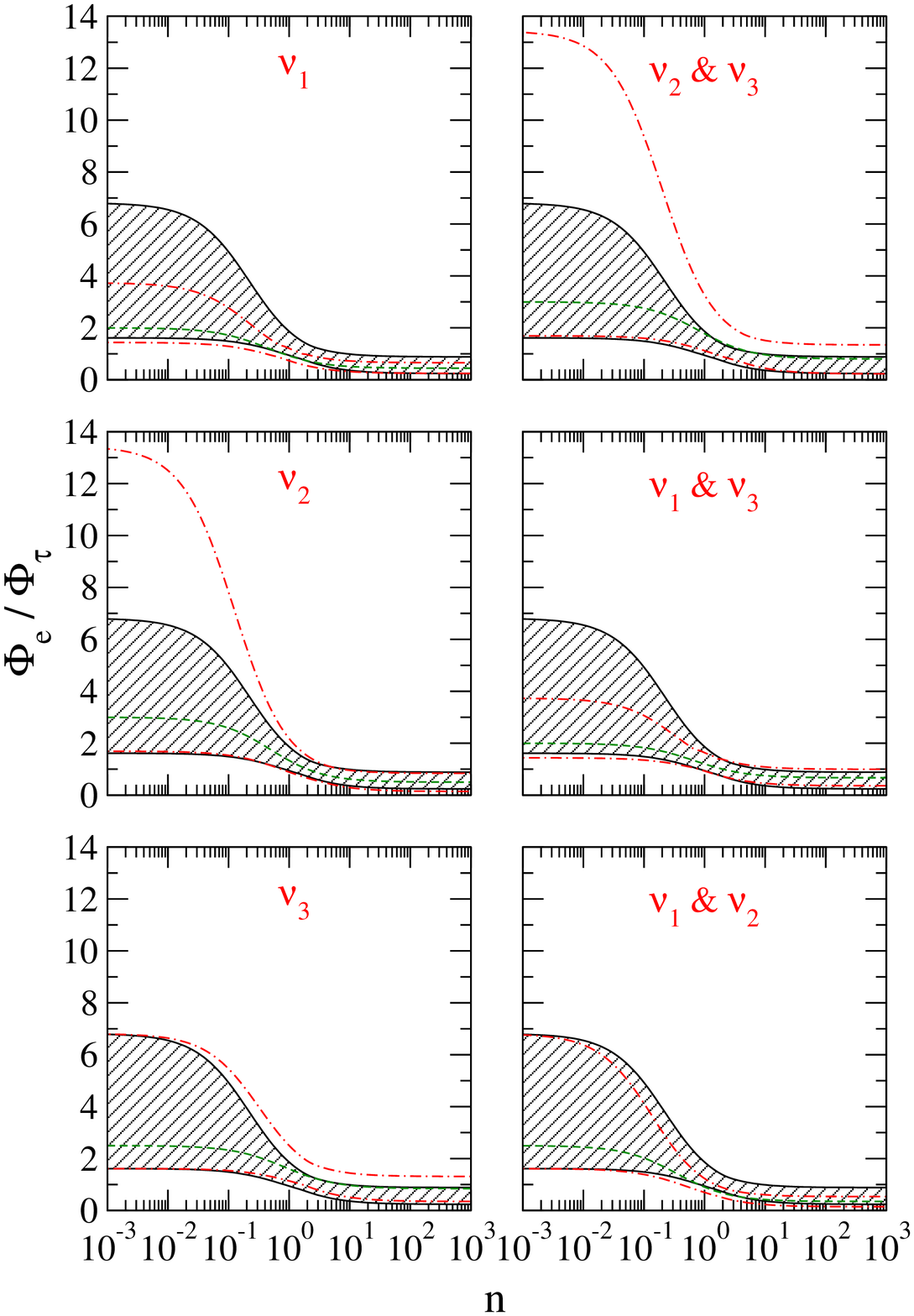}
 \caption{Same as Fig.~\ref{fig:ratio_1enmu} for the observable flux ratio $\ratioetau$.}
\label{fig:tauratio_1enmu}
\end{figure}

\begin{figure}[htp]
 \centering
 \includegraphics[width=0.8\textwidth]{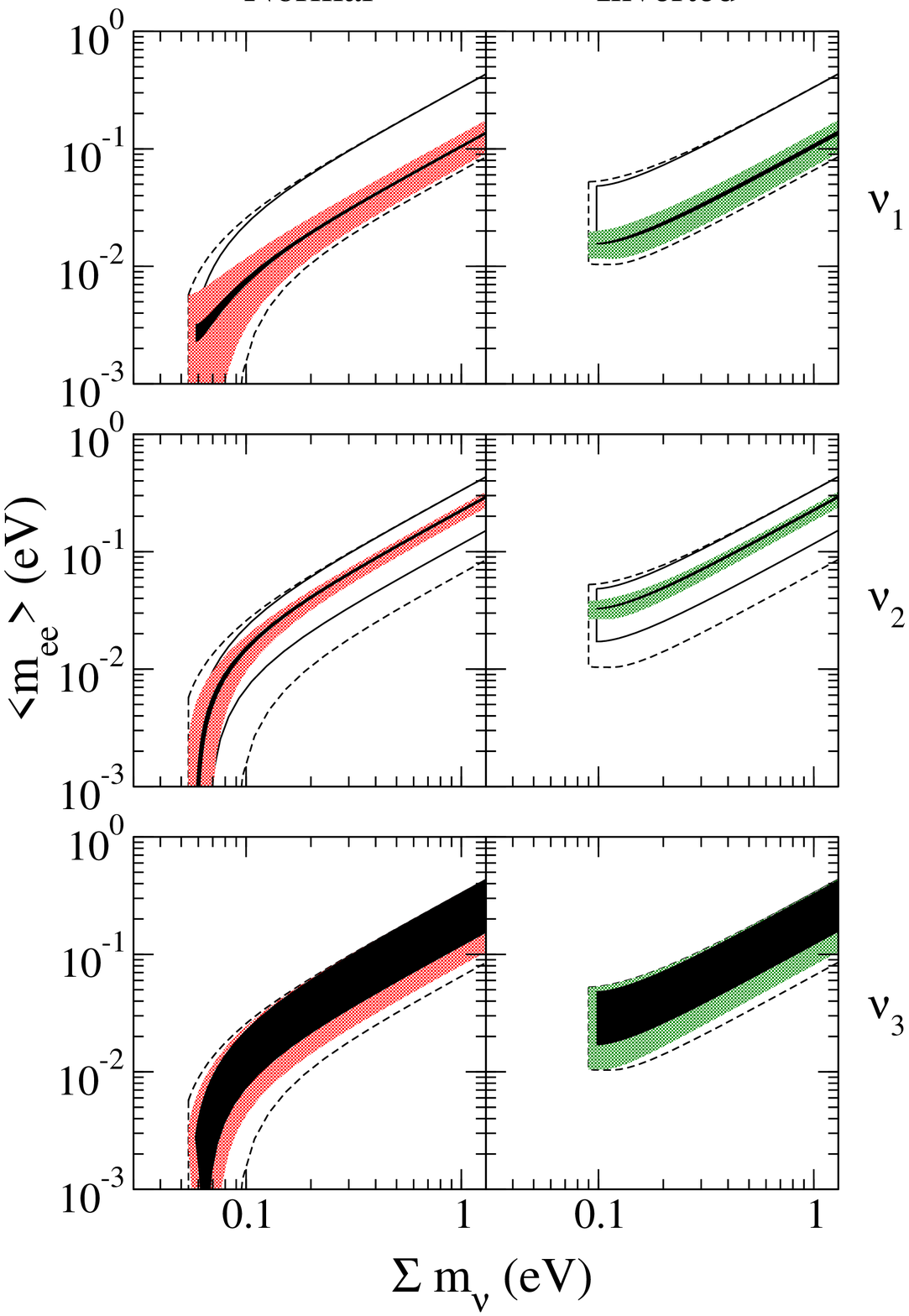}
 \caption{Allowed regions in the $\mee-\sumnu$ plane for the three different cases of one pseudo-Dirac neutrino (indicated on the right of each row). The black regions are for exact TBM, and the light red (green) shaded regions correspond to the $3\sigma$ ranges of the oscillation parameters for normal (inverted) ordering. The solid (dashed) lines indicate the best-fit ($3\sigma$) allowed regions in the standard three-neutrino scenario.}
\label{fig:mee_sum_1dirac}
\end{figure}

\begin{figure}[htp]
 \centering
 \includegraphics[width=0.8\textwidth]{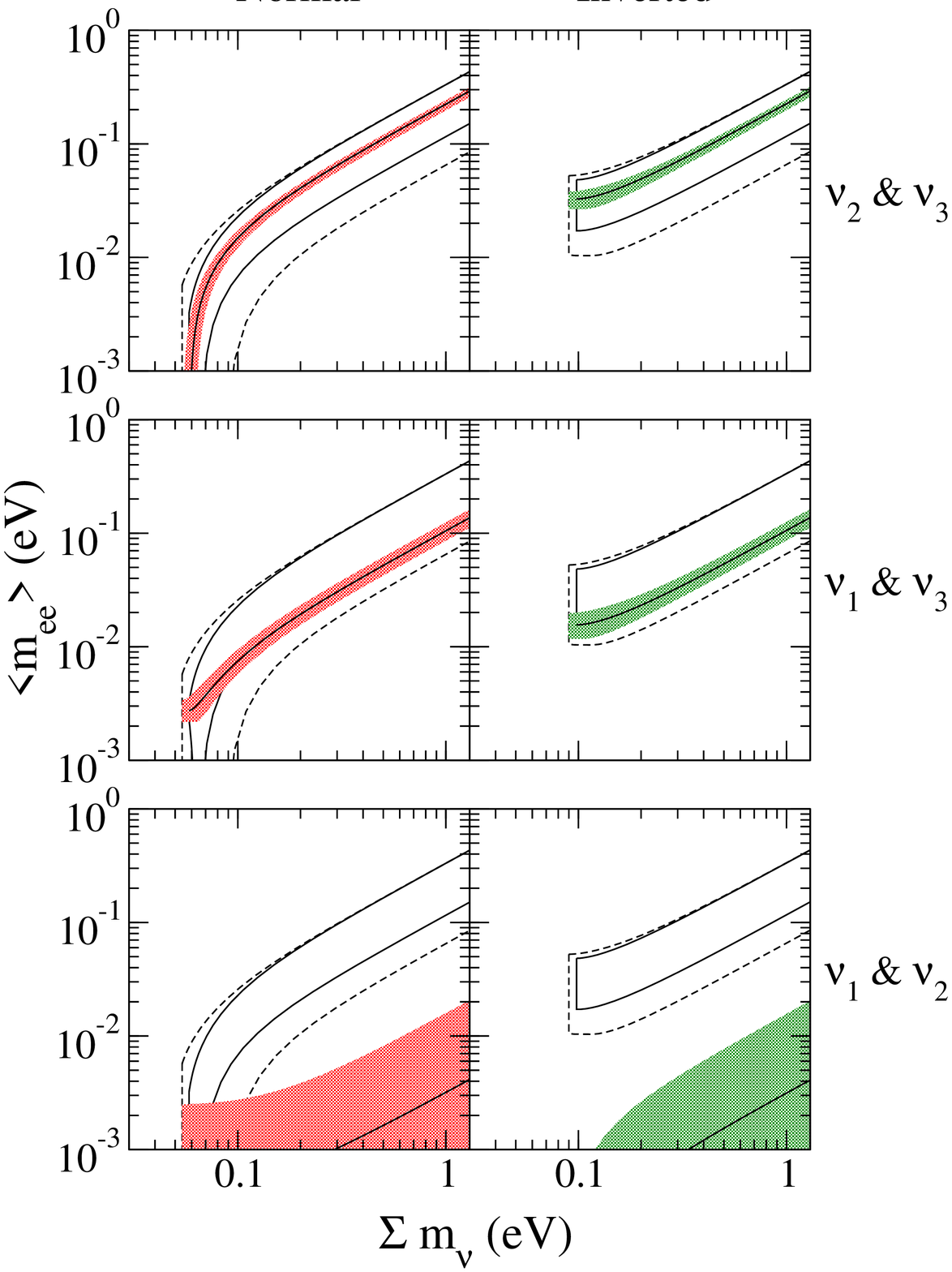}
 \caption{Same as Fig.~\ref{fig:mee_sum_1dirac} for the three different cases of two pseudo-Dirac neutrinos (indicated on the right of each row).}
\label{fig:mee_sum_1majorana}
\end{figure}

In conclusion, we have studied two different ways to test the
bimodal (schizophrenic) neutrino hypothesis that one or two of the
neutrino mass eigenstates are Dirac particles and the others
Majorana. There are in total six nontrivial possible combinations,
and we have performed a mostly phenomenological analysis of these
scenarios. We noted that (i) flux ratios of extra-galactic high
energy neutrinos, and (ii) the effective mass for neutrinoless
double beta decay are sensitive to the different possibilities,
showing nonstandard behavior in many cases.
Figs.~\ref{fig:ratio_1e2mu} to \ref{fig:mee_sum_1majorana}
summarize our results. In brief, we found that flux ratios can differ significantly from
their standard values and the effective mass can either lie only
in certain regions or even completely outside of its standard
parameter space. The examples given show that the many different
experimental signatures provide good tests of whether neutrino
masses have the bimodal (or pseudo-Dirac) character. We have also
discussed simple beyond the standard model scenarios in which
such bimodal features can arise. Evidence for
bimodal nature of neutrino mass will require major changes in our
thinking about the physics of neutrino mass. 
Indeed, the field of neutrino physics has provided many surprising
results in the past, and the question of neutrino mass origin is far
from settled. If the hypothesis of bimodal neutrinos is supported by
the experiments outlined here, it will not only provide a major
departure from our current thinking about the nature of neutrino
masses but also its theoretical origin from physics beyond the
standard model. As such it will have major impact on the physics at
and beyond the TeV scale. 
   
\begin{center}
{\it \bf Acknowledgements}
\end{center}

JB and WR are supported by the ERC under the Starting Grant MANITOP
and by the DFG in the project RO 2516/4-1 as well as in the Transregio
27. R.N.M.~is supported by the NSF under Grant No.~PHY-0968854.



\end{document}